\documentstyle[epsfig]{l-aa} 
\def\aeta{A\&A }

\def\apj{ApJ }
\def\apjs{ApJS }

\def\mn{MNRAS }

\def\apjl{ApJL \rm}

\begin{document}

\voffset -2.5truecm

\thesaurus{12.12.1;11.09.3;11.17.1}

\title{Evolution of the Lyman$\alpha$ forest from high to low redshift}
\author{R\"udiger Riediger$^1$, Patrick Petitjean$^{2,3}$, and Jan~P.\ M\"ucket$^1$}
\institute{
$^1$Astrophysikalisches Institut Potsdam, An der
Sternwarte 16, D-14482  Potsdam, Germany\\
$^2$Institut d'Astrophysique de Paris -- CNRS, 98bis Boulevard Arago,
F-75014 Paris, France\\
$^3$UA CNRS 173- DAEC, Observatoire de Paris-Meudon, F-92195 Meudon
Principal Cedex, France}
\date{ }
\offprints{R.\ Riediger}
\maketitle
\markboth{}{}
\begin{abstract}
We study the evolution with redshift, from \mbox{$z\sim5$} to \mbox{$z=0$}, 
of the Lyman$\alpha$ forest in a CDM model using numerical simulations 
including collisionless particles only. The baryonic component is assumed 
to follow the dark matter distribution. \par
We distinguish between two populations of particles: Population $P_{\rm s}$
traces the filamentary structures of the dark matter, evolves slowly with 
redshift and, for
\mbox{$N(\mbox{H{\sc i}}) \ga 10^{14}\mbox{ cm$^{-2}$}$}, dominates the 
number density of lines at \mbox{$z<3$}; 
most of population $P_{\rm u}$ is located in underdense regions
and for the same column densities, disappears rapidly at high redshift. \par
We generate synthetic spectra from the simulation and show that the
redshift evolution of the Lyman$\alpha$ forest (decrement,
$N(\mbox{H{\sc i}})$ distribution) is well reproduced over the whole
redshift range 
for \mbox{$\Omega_{\rm b}h^2 \sim 0.0125$} and 
\mbox{$J_{-21} \approx 0.1$} at \mbox{$z \sim 3$} where $J_{-21}$ 
is the UV background flux intensity in units of
\mbox{$10^{-21}$ erg cm$^{-2}$ s$^{-1}$ Hz$^{-1}$ sr$^{-1}$}.\par
The total number of lines with
\mbox{$N(\mbox{H{\sc i}}) \ga 10^{12}\mbox{ cm$^{-2}$}$} remains 
approximately constant from \mbox{$z\sim 4$} to \mbox{$z=1$}.
At \mbox{$z\sim 0$}, the number density of lines per unit redshift with 
\mbox{$\log N(\mbox{H{\sc i}}) >12, 13, 14$} is of the order of 400, 100, and 
20 respectively. Therefore, at low redshift,
if most of the strong (\mbox{$w_{\rm r}>0.3 \mbox{ \AA}$}) lines are
expected to be associated
with galaxies, the bulk of the Lyman$\alpha$ forest however 
should have lower equivalent width and should not be tightly correlated
with galaxies.
\end{abstract}
\section{Introduction}\label{intr}
Recent high spectral resolution and high ${\rm S}/{\rm N}$ ratio 
observations of the
Lyman$\alpha$ forest have shed some light on previous discussions about
the physical properties of the gas. 

There is a cut-off in the Doppler parameter distribution with only
a very small fraction of narrow lines (\mbox{$b<15 \mbox{ km s$^{-1}$}$})
which cannot be identified with metal-lines (Rauch et al.\ 1992). 
The large $b$ values could
reflect blending of several components with similar properties 
(Hu et al.\ 1995), in which case the gas would have a quite
homogeneous temperature (\mbox{$T \approx 24\,000 \mbox{ K}$}), and turbulent
motions inside one 
cloud would be small. The H{\sc i} column density distribution 
is well fitted by a power-law
\mbox{${\rm d}n/{\rm d}N \propto N^{-\beta}$} with
\mbox{$\beta \approx 1.5$} in the range \mbox{$13.5 < \log N(\mbox{H{\sc i}}) < 15.5$}
(Hu et al.\ 1995, Lu et al.\ 1996). For 
\mbox{$\log N(\mbox{H{\sc i}}) \ga 15$}, there is a deficit of lines 
(Petitjean et al.\ 1993), compared to the power-law function. 
The amplitude of the deficit seems to evolve with redshift, being larger at 
\mbox{$z\sim 2.5$} than at \mbox{$z\sim 3.5$} (Kim et al.\ 1997). 

For
\mbox{$\log N(\mbox{H{\sc i}}) < 13.5$}, blending effects depress the observed distribution.
Synthetic spectra generated from a random realisation out of
an artificial population of clouds show that the underlying column density
distribution has no turn over at low column density. The correction
however can be very large. At \mbox{$\log N(\mbox{H{\sc i}}) \approx 12.4$}, 6\%~of the lines
are recovered at \mbox{$z\sim 3.7$} (Lu et al.\ 1996, Kim et al.\ 1997).
The way the correction is done may not be the unique solution. 
In particular, if the Lyman$\alpha$ complexes have large dimensions,
the redshift range corresponding to the Hubble flow and velocity
structure of the cloud along the radial dimension must be avoided.
The large dimensions (\mbox{$100 - 300 \mbox{ kpc}$},
Dinshaw et al.\ 1995, Smette et al.\ 1995, Crotts \& Fang 1997)
recently derived from observations of lines of sight
towards quasars projected on the sky at small separations may
support this remark. The existence of a turn-over in the column density
is thus still a matter of debate; the question being related to the effective
size of the complexes. More data are needed to investigate this question.\par
One of the most important issue to clarify towards understanding nature 
of the Lyman$\alpha$ forest is the relation between the Lyman$\alpha$ forest
and galaxies. 
Although conclusions are uncertain, it seems that at least the strongest
lines in the Lyman$\alpha$ forest at low redshift are
anyhow associated with galaxies (Lanzetta et al.\ 1995,
Le Brun et al.\ 1996).
This is not really surprizing since if there is gas in the intergalactic medium
(and it seems there is), we can expect the density of this gas to be
higher in the vicinity of the galactic potential wells. 
The case for the weak lines to be associated with galaxies is less clear.
Indeed observations of the line of sight to 3C273 that are the most
sensitive to the presence of weak lines (Morris et al.\ 1991,
Bahcall et al.\ 1991) indicate the presence
of a large number of these lines and no clear association with galaxies
is seen (Morris et al.\ 1993, see also Bowen et al.\ 1997). In addition,
using a high ${\rm S}/{\rm N}$ ratio spectrum of H1821+643 
(\mbox{$z_{\rm em}=0.297$}), Tripp et al.\ (1997) have found that,
the number per
unit redshift of lines with \mbox{$w_{\rm r}>50 \mbox{ m\AA}$} is $112\pm 21$.
Moreover, Stocke et al.\ (1995)
have detected 
weak absorption lines located in regions devoided of galaxies.\par
Recent observations have shown that at redshift \mbox{$z \sim 3$}, C{\sc iv} is found 
in 90\% of the clouds with
\mbox{$N(\mbox{H{\sc i}}) > 10^{15} \mbox{ cm$^{-2}$}$} and in 
about 50\% of the clouds with 
\mbox{$3\times 10^{14} \mbox{ cm$^{-2}$} < N(\mbox{H{\sc i}}) < 10^{15} \mbox{ cm$^{-2}$}$}
(Songaila \& Cowie 1996, Cowie et al.\ 1995).
Several components are seen in most of these
weak systems; thus the two-point correlation function shows a signal on scale
smaller than \mbox{$200 \mbox{ km s$^{-1}$}$}. On this basis,
Fern\'andez-Soto et al.\ (1996)
show that the observed clustering is broadly compatible with that
expected for galaxies.
It can be argued however that the signal detected in the correlation function 
is not related to clustering of galaxies but instead reflects
the velocity structure of the Lyman$\alpha$ gas inside large complexes.
Although such clustering is observed as well for metal line systems
(Petitjean \& Bergeron 1994, see Cristiani et al.\ 1996), this does not mean 
that both Lyman$\alpha$ complexes and metal line systems are identical in 
nature since Lyman$\alpha$ complexes have lower abundances 
(Hellsten et al.\ 1997) and larger dimensions.\par
Indeed a more attractive picture arises from simulations showing that
the Lyman$\alpha$ absorption line properties can be understood if the
gas traces the gravitational potential of the dark matter
(Cen et al.\ 1994, Petitjean et al.\ 1995,
M\"ucket et al.\ 1996, Hernquist et al.\ 1996, 
Miralda-Escud\'e et al.\ 1996, Zhang et al.\ 1996, Bi \& Davidsen 1996). 
In this picture, part of the gas is located inside filaments
where star formation can occur very early in small halos that subsequently
merge to build-up a galaxy (Haehnelt et al.\ 1996). 
It is thus not surprizing to observe metal lines from this gas.
The remaining part of the gas
either is loosely associated with the filaments and 
has \mbox{$N(\mbox{H{\sc i}}) \ga 10^{14} \mbox{ cm$^{-2}$}$}
or is located in the underdense regions and has 
\mbox{$N(\mbox{H{\sc i}}) \la 10^{14} \mbox{ cm$^{-2}$}$}.
Most of the Lyman$\alpha$ forest arises in this gas and
it is still to be demonstrated that it contains metals.\par
To clarify these issues, 
it is important to follow the evolution of the 
Lyman$\alpha$ gas over a wide redshift interval up to the 
current epoch (\mbox{$z = 0$}) 
distinguishing between gas 
closely associated with the filamentary structures of the 
dark matter and the dilute gas 
mainly located in the underdense regions. The problem 
which the true hydrodynamic simulations are still confronted with is the 
huge amount of computer time needed to model a whole line of sight. 
The goal of our paper is then twofold: to investigate the 
time-dependence of the Lyman$\alpha$ forest along the
whole line of sight, and to investigate the relative contributions
of the two different gas distributions to the forest characteristics.
In Section 2. we discuss the main assumptions of the model and
the limits of applicability.
Results and predictions of the model are presented in Section 3.
and conclusions are drawn in Section 4.  
\section{Model and simulations}\label{simul}
%
%
%
\begin{figure}
\begin{minipage}[b]{.7\linewidth}
\epsfig{file= 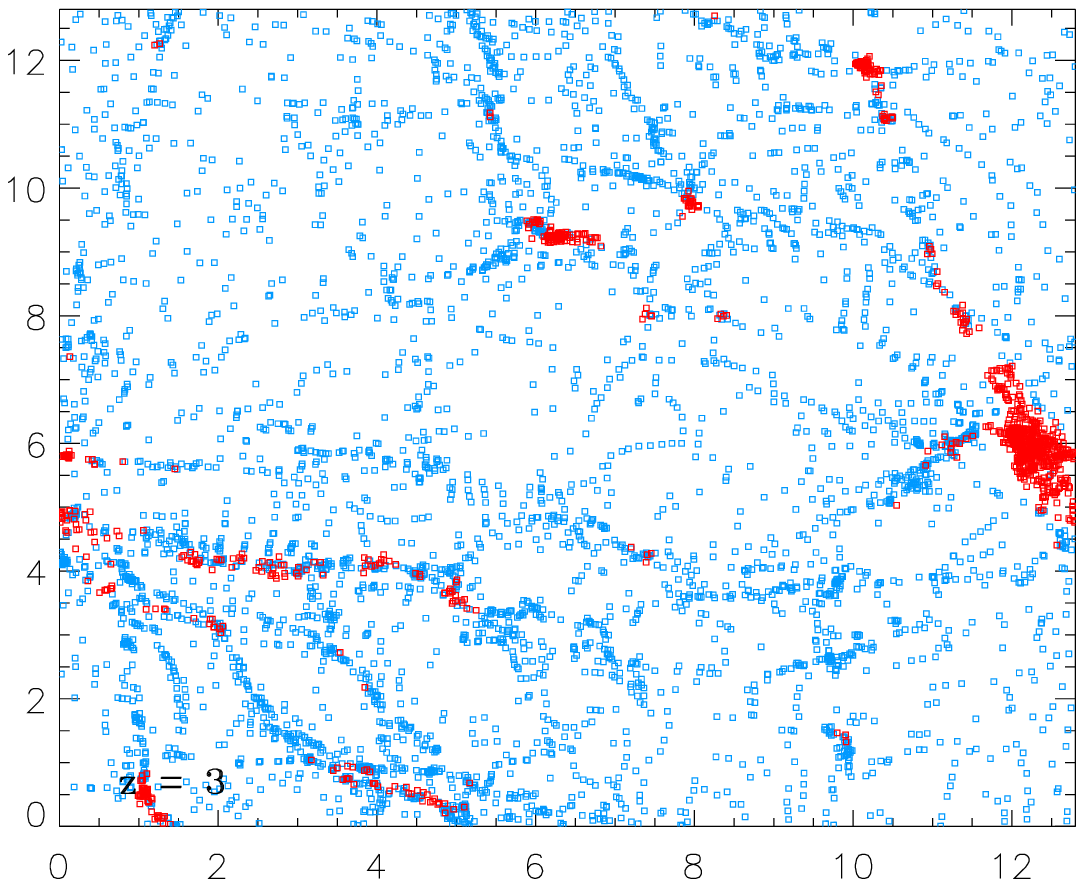 ,scale=1., height=7cm,width=8cm}
\end{minipage}

\begin{minipage}[b]{.7\linewidth}
\epsfig{file= 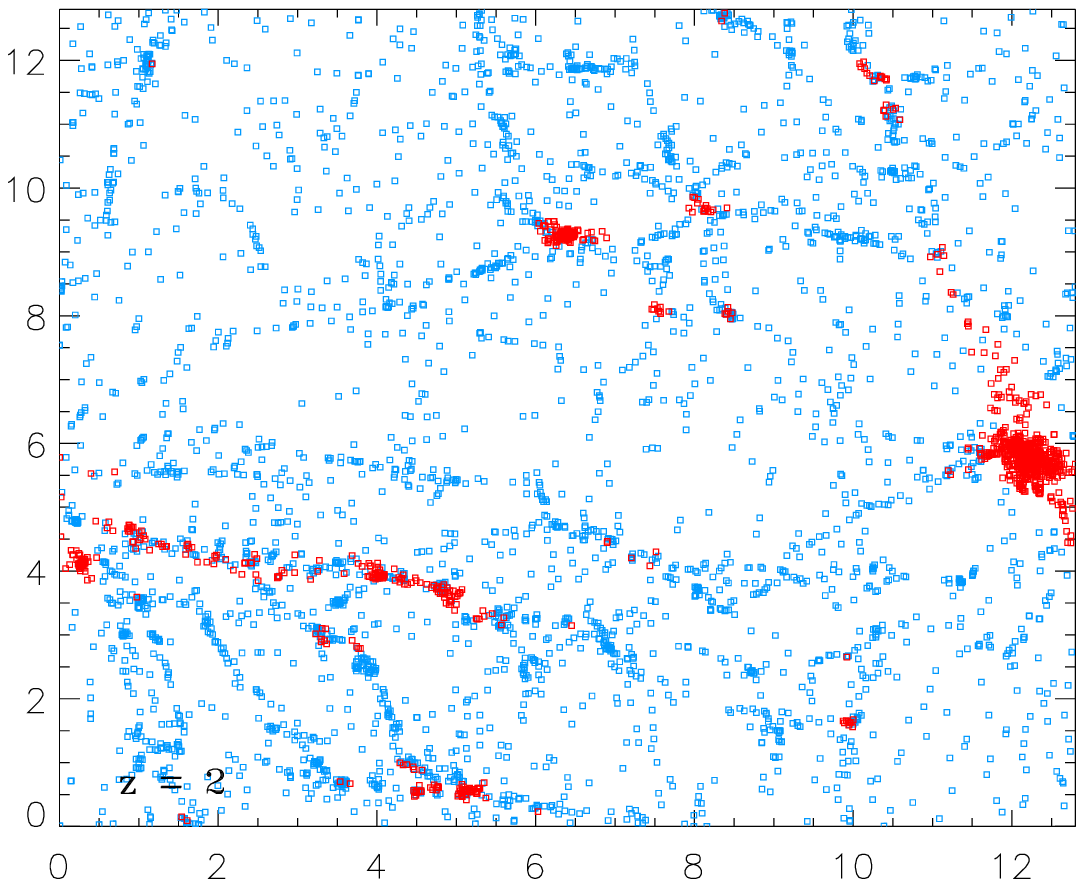 ,scale=1., height=7cm,width=8cm}
\end{minipage}

\begin{minipage}[b]{.7\linewidth}
\epsfig{file= 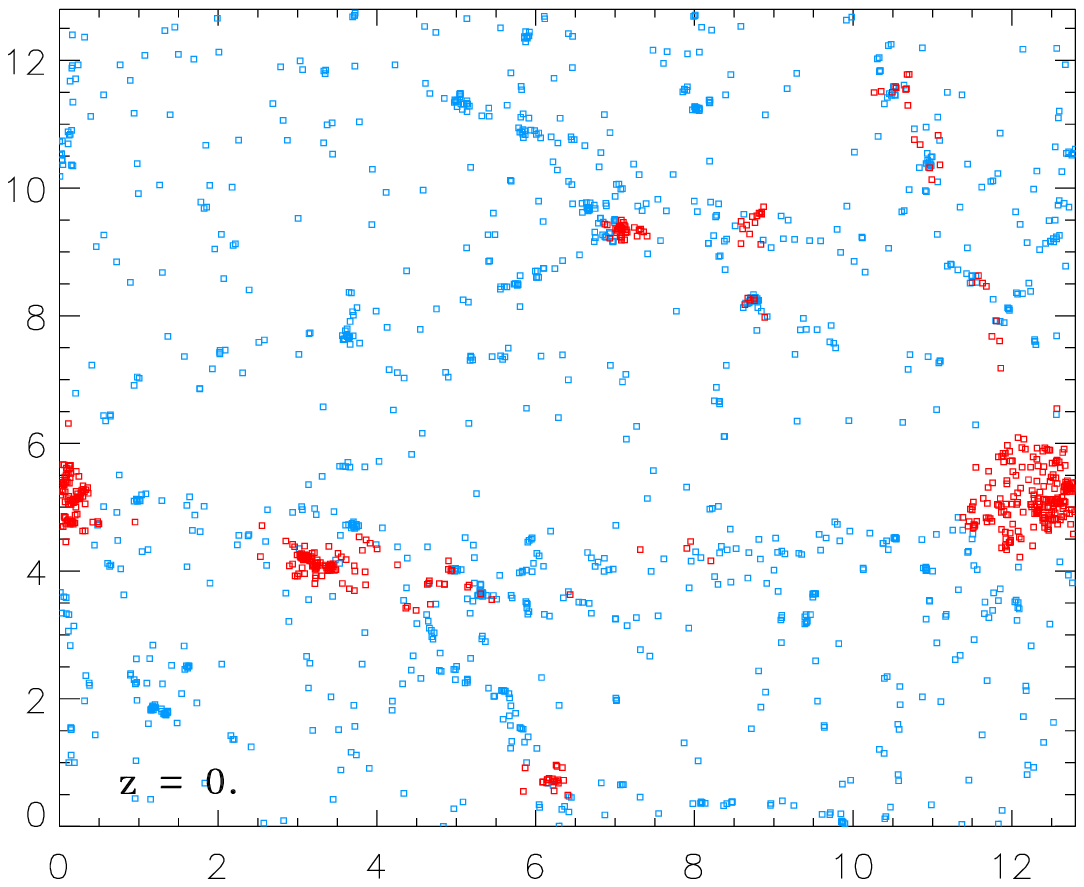 ,scale=1., height=7cm,width=8cm}
\end{minipage}
\caption[]{``Unshocked" (open squares) and ``shocked" (filled squares) 
clouds in a
slice of 50~kpc thickness of a simulation (12.8~Mpc)$^3$
box at different redshifts: \mbox{$z=3$} (top), \mbox{$z=2$} (middle),
\mbox{$z=0$} (bottom); the size of the symbols is comparable to the
resolution.}
\label{3fig}
\end{figure}
We have used the particle-mesh code described in detail by Kates et al.\ (1991),
modified as described in our previous paper (M\"ucket et al.\ 1996). 
A particle simulation is used in order to investigate
the evolution of the Lyman$\alpha$ forest over a wide redshift range
using two main assumptions:

First, the Lyman$\alpha$ gas traces the shallower potential wells of the 
dark matter distribution. In previous papers 
(Petitjean et al.\ 1995, M\"ucket et al.\ 1996) we have already 
discussed in detail the applicability of this approximation as well as its 
consequences: the gas with column density smaller than
\mbox{$10^{17} \mbox{ cm$^{-2}$}$} closely follows the dark matter distribution. 
This assumption has been shown to be valid
for the low-density regime characteristic of
the Lyman$\alpha$ forest by detailed hydro-simulations (see  
Miralda-Escud\'e et al.\ 1996,
Hernquist et al.\ 1996, Yepes et al.\ 1997, and also Gnedin et al.\ 1997,
Hui et al.\ 1997). 

Secondly, a procedure to determine the gas temperature has to 
be introduced. 
We follow Kates et al.\ (1991) where it is assumed that
energy is acquired by the gas, in addition to photo-ionization, when 
shell crossing takes place in the dark matter dynamics. Particles
that have acquired this way enough energy to increase the temperature 
above \mbox{$T>T_{\rm th}=20\,000\mbox{ K}$} are called 'shocked' (Population
$P_{\rm s}$).
This temperature corresponds to the lower limit of the range of equilibrium 
temperatures found for different densities and UV flux intensities in highly 
ionized metal-poor gas.
Other particles are called 'unshocked' (Population $P_{\rm u}$). 
The thermal history is followed as described in M\"ucket et al.\ (1996)
taking into account adiabatic effects, radiative cooling processes, 
Compton cooling and heating. 
As discussed in the previous paper, this is a reasonable treatment as far
as the Lyman$\alpha$ forest is concerned.
Clearly our approximation fails for column densities 
\mbox{$N(\mbox{H{\sc i}})>10^{17} \mbox{ cm$^{-2}$}$}. Therefore a
description of the distribution and state of the gas in dense collapsing 
regions would lead to uncertain results.

To first order, the distinction between 'shocked' and 'unshocked' gas 
in the code
is independent of resolution since shocks should be supressed by
thermal pressure in population $P_{\rm u}$.
The approximation is valid if the grid size is smaller than the Jeans length. 
According to M\"ucket et al.\ (1996), 
this requires 
\mbox{$2l_{\rm c} \le 1.2 \cdot (T_4/(1+z_i))^{1/2} \mbox{ Mpc}$}
where $l_{\rm c}$ denotes the comoving length of a cell. 
At \mbox{$z = 5$} we obtain 
\mbox{$l_{\rm c} \approx 240\mbox{ Mpc}$} which is
fairly well realized in our simulations. The gas constituent of
each particle can thus be considered as a non-collapsing diffuse gas element.

The gas is considered in thermal equilibrium. 
The assumption is no longer valid at very 
low densities. Our simulation is limited to the cosmic background 
density however
and thus to \mbox{$N(\mbox{H{\sc i}})>10^{12} \mbox{ cm$^{-2}$}$} where 
deviation from equilibrium is minimized. 

The simulations use $128^3$ particles on a $256^3$ grid
and were carried out using a box size of 12.8~Mpc, which corresponds to
a co-moving cell size of 50~kpc.  
The baryonic mass is assumed to be proportional to the dark matter mass
inside a cell. We adopt a value for the Hubble parameter
\mbox{$H_0 = 50\mbox{ km Mpc$^{-1}$ s$^{-1}$}$} and 
\mbox{$\Omega_{\rm b} = 0.05$} throughout.

The intensity of the photo-ionizing UV background flux,
assumed to be homogeneous and isotropic inside the simulation box,
is computed in the course of the simulation. 
The ionizing spectrum is
modeled as \mbox{$J_{\nu} \propto J_0\,\nu^{-1}$} where
\mbox{$J_0= f(z) = J_{-21} \cdot 10^{-21} 
\mbox{ erg cm$^{-2}$ s$^{-1}$ Hz$^{-1}$ sr$^{-1}$}$}
is the ionizing flux at 13.6~eV.
The variation of the flux intensity with redshift $z$ is related both to  
the rate \mbox{$\Delta m(T_4<0.5;z)$} at which the baryonic
material cools below \mbox{$T_4=0.5$}
(with \mbox{$T = T_4 \cdot 10^4 \mbox{ K}$}) in the simulation and to
the expansion of the Universe:
\begin{eqnarray}
	f(z) & = & C_{\rm cool}\, \Delta m(T_4<0.5; z) \nonumber\\
	& &{} + f(z+\Delta z) \left(\frac{1+z}{1+z+\Delta z}\right)^4,
	\label{flux1} 
\end{eqnarray}
where $C_{\rm cool}$ is a factor of proportionality. 
To avoid overcooling (e.g.\ Blanchard et al.\ 1992), we assume that 
the cool gas is transformed into stars with an efficiency $\varepsilon$ 
of about 8\%. 
The remainder of the gas is reheated to temperatures above 
50\,000~K. The characteristic time scale for such processes 
is of order \mbox{$t_\ast \approx 10^8 \mbox{ years}$}. This procedure  
provides therefore results that are independent of the time step.
Whereas the time dependence of the UV flux is almost not affected by
the value of $t_\ast$ within a reasonable range, the value 
of the efficiency parameter $\varepsilon$ has considerable influence on the 
flux intensity 
at small redshifts (see also Miralda-Escud\'e et al.\ 1996): 
in this scenario it determines the amount of gas still available at
any time for star formation.

The dynamics of structure formation is highly
sensitive to the initial fluctuation spectrum. Since we are working here
on scales mainly determined by the $k^{-3}$ tail of the CDM spectrum, only
the effective amplitude of the spectrum within the $k$-range of
interest is of importance. 
In particular it 
affects the redshift evolution of the computed
UV flux intensity. We have used this dependence to constrain the power of the
initial density fluctuations in the simulation in order to be able to
reproduce the observed evolution (see Section~\ref{decrement}).
The resulting effective power is 10\% larger than what is expected from a
COBE spectrum normalized at large scales.  

During the simulation we distinguish between the two different populations of
clouds defined above namely: population $P_{\rm s}$ corresponds to 
particles for which shock-heating has been important at some time of the 
thermal history of the particle and population $P_{\rm u}$ to particles for 
which photo-ionization is the dominant heating process all the time.
Population $P_{\rm s}$ particles are mostly found in big halos and elongated
filamentary structures (regions of enhanced density).
The remainder of the gas (population $P_{\rm u}$) is present in the 
surroundings of
the structures formed by shocked particles but mostly in the voids
delineated by these structures. \par
The spatial distribution of both populations of particles at different 
redshifts (\mbox{$z = 3, 2 \mbox{, and } 0$}) in a slice of dimensions 
\mbox{$12.8 \times 12.8 \times 0.05 \mbox{ Mpc$^3$}$}
(comoving) is illustrated in 
Fig.~\ref{3fig}. In this figure all particles in the slice are shown regardless
of their neutral hydrogen content. 
The fraction of particles belonging to population $P_{\rm u}$
is changing with time: 77\% at \mbox{$z=3$}, 67\% at \mbox{$z=2$}, 40\% at
\mbox{$z=0.5$} and 25\% at \mbox{$z=0$}.
\section{Results}\label{res}
\subsection{Synthetic spectra}\label{spectra}
The modeling of the cloud distribution along a full line of sight up to a 
fictitious QSO at redshift \mbox{$z = 5$} is described in detail in
M\"ucket et al.\ (1996).
We use the simulation 
to synthesize spectra along a line of sight. Two examples of such 
spectra 
are given in Fig.~\ref{spectrum}. 
The upper and lower panels show the Lyman$\alpha$ forest at 
\mbox{$z\sim3$} and 0.3 respectively with 
resolution and ${\rm S}/{\rm N}$ ratio similar to
Keck/HIRES and future HST/STIS observations respectively (see 
Section~\ref{evolution}).
The wavelength ranges are chosen to correspond to the same comoving distance.

To analyse the data, we have developped an automatic line fitting procedure 
within the ESO data reduction package MIDAS taking advantage of 
the context FITLYMAN. The procedure selects
regions of the spectrum between two wavelengths at which the normalized
continuum is unity. Within such a region, it determines the position of
the minima and starts by placing one component at each minimum,
considering the central wavelengths as fixed parameters. Column densities and 
Doppler parameters are determined minimizing the reduced $\chi^2$. 
Components are then introduced where the difference between the model 
and the data is at maximum. A new fit is performed relaxing the positions. 
A detailed investigation of the ability of automatic line profile fitting
to derive reliable information on the underlying population
will be presented elsewhere.

\subsection{The average Lyman$\alpha$ decrement}\label{decrement}
\begin{figure} 
\epsfig{file=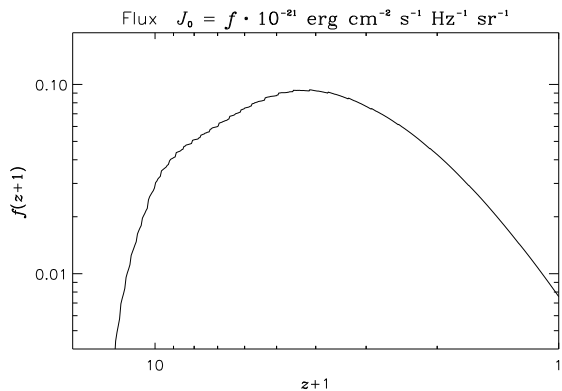 ,scale=1., width=8cm}
\caption[]{Evolution of the UV-background intensity from the simulation}
\label{flux}
\end{figure} 
\begin{figure} 
\epsfig{file=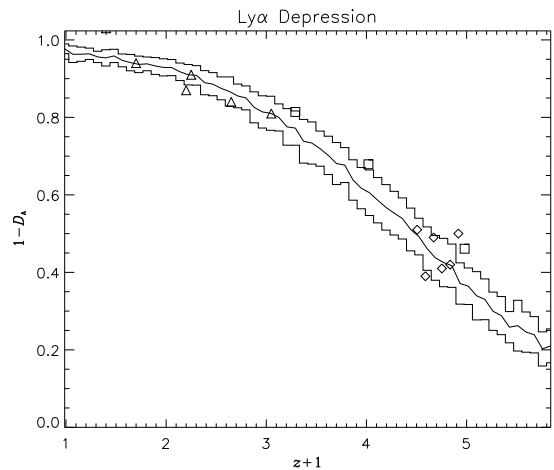,scale=1., width=8cm}
\caption[]{Lyman$\alpha$ decrement versus redshift.
Observed data are from: diamonds, Lu et al.\ (1996);
triangles, a compilation by Jenkins \& Ostriker\ (1991) at low redshifts; 
squares: Rauch et al.\ (1997)}
\label{da}
\end{figure}  
\begin{figure*} 
\epsfig{file=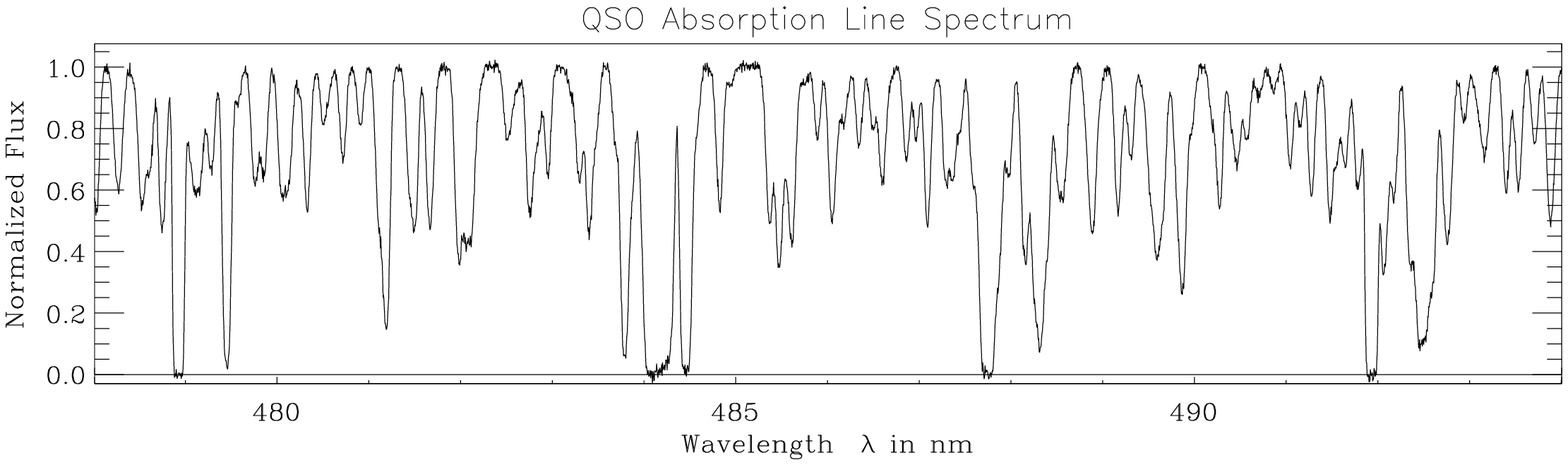,scale=1., height=5cm,width=18cm}
\epsfig{file=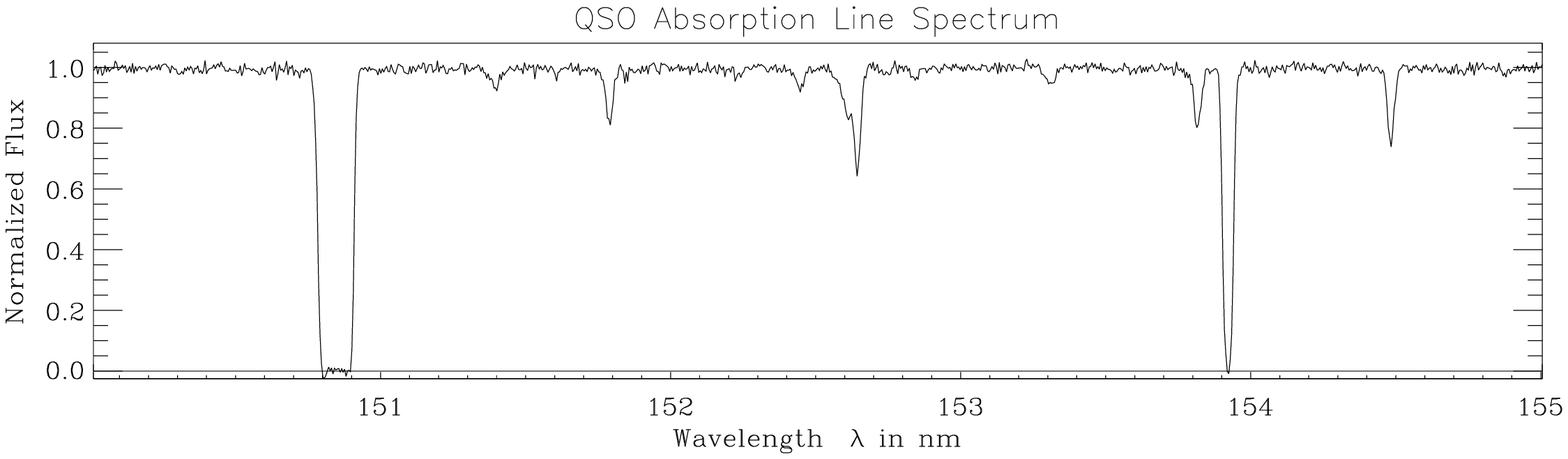,scale=1., height=5cm,width=18cm}
\caption[]{Portion of a synthetic spectrum derived from 
the simulation at high redshift $z\sim3$ (top)
and low redshift $z\sim0.3$ (bottom).}
\label{spectrum}
\end{figure*}
Due to blending effects, comparison between simulations and observations 
is not simple. A first test of the model is that 
the evolution of the average Lyman$\alpha$ 
decrement should be reproduced. As emphasized by Miralda-Escud\'e et 
al.\ (1996), the decrement
depends on the mean H{\sc i} density which is directly related to
\mbox{$J_0^{-1}(\Omega_{\rm b} h^2)^2$} where $\Omega_{\rm b}$ is
the baryon density and
\mbox{$J_0=J_{-21} \cdot 10^{-21} \mbox{ erg cm$^{-2}$ s$^{-1}$ Hz$^{-1}$ sr$^{-1}$}$}
is the ionizing flux at 13.6~eV. 
In our simulations, the baryon density is given a value 0.05 
(\mbox{$\Omega_{\rm b} h^2 = 0.0125 $}) and
the evolution with redshift of the ionizing flux is computed assuming
that its variation is related to the amount of gas that collapses in the 
simulation at any time (see Section~\ref{simul}). The only free parameter 
is thus the normalization of \mbox{$J_{-21}(z)$}.
A value of \mbox{$J_{-21}(z_0)=0.1$} at \mbox{$z_0 \sim 3$} 
(see Fig.~\ref{flux}) fits the decrement evolution quite well 
(see Fig.~\ref{da}).   

Our result for the intensity of the UV-background at \mbox{$z\sim3$}
is consistent with the findings by Hernquist et al.\ (1996) and
Miralda-Escud\'e et al.\ (1996) using hydro simulations over a much smaller
redshift range. A lower limit \mbox{$J_{-21}(z_0)=0.16$} is obtained
however by Rauch et al.\ (1997) from the observed number of sources and 
absorbers
of ionizing photons. A reason for this discrepancy could be that we
have adopted a constant spectral shape for the UV
spectrum \mbox{$\propto \nu^{-1}$}. 
Since the 
ionization equilibrium is determined by the ionization parameter
that is the ratio of the density of ionizing photons to the hydrogen density,
a steeper spectrum or a break at 54.4~eV (e.g Haardt \& Madau, 1996)  
would have given a larger 
normalization by a factor of about 2.

Determinations of the UV background intensity 
from observations of the pro\-xim\-ity-ef\-fect however 
give much larger values, 
\mbox{$J_{-21}(z) \approx 0.5$} in the redshift range
\mbox{$1.7 < z < 3.8$} (Bechtold 1994; Bajtlik et al.\ 1988; Lu et al.\ 1991,
Giallongo et al.\ 1996, Cooke et al.\ 1996)
and \mbox{$J_{-21}(z) \approx 0.2$} at \mbox{$z \sim 4.2$}
(Williger et al.\ 1994).
This is up to five times larger than what we obtain for the corresponding
redshifts. 
One reason for this discrepancy could be that the
basic assumption made when applying the
proximity-effect technique (Bajtlik et al.\ 1988) is not valid.
Indeed it is assumed that 
in the absence of the QSO ionizing flux, 
the number density of the lines in the vicinity of the quasar
would evolve according to an 
extrapolation of the power law  
\mbox{${\rm d}n/{\rm d}z \propto (1 + z)^\gamma $} found far away from
the quasar. It seems probable
however that QSOs are located in regions of enhanced density. 
In that case a strong increase of the number of lines in the vicinity
of quasars should be expected. If true, the technique would systematically  
overestimate the UV background intensity.
This has been mentioned previously by Bechtold (1994) and
Williger et al.\ (1994).

At low redshift, Kulkarni \& Fall (1993) derive 
\mbox{$J_{-21}(z=0) \approx 0.002-0.04$} from a study of the proximity effect
using HST data. Donahue et al.\ (1995) give 
\mbox{$J_{-21}(z=0) < 0.03$} as an upper limit on the local ionizing 
background based on a search for extended Lyman$\alpha$ emission in nearby
extragalactic H{\sc i} clouds.
\begin{figure} 
\epsfig{file=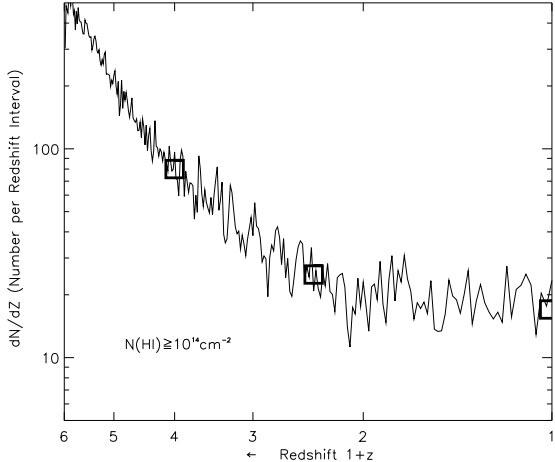,scale=1., width=8cm}
\caption[]{Number density ${\rm d}n/{\rm d}z$ of clouds with
column density \mbox{$\log N(\mbox{H{\sc i}}) > 14$} versus redshift $z$}
\label{dndz_all}
\end{figure} 
\begin{figure}
\epsfig{file=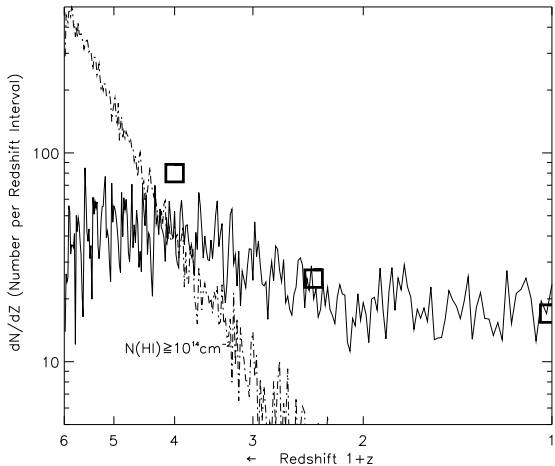,scale=1., width=8cm}
\caption[]{As Fig.~\ref{dndz_all}. The number density of
lines drawn from populations $P_{\rm s}$ and $P_{\rm u}$ are plotted as 
full and dash-dotted lines respectively}
\label{dndz_sep}
\end{figure} 
\subsection{Evolution of the line number density}\label{evolution}
Fitting the Lyman$\alpha$ forest is ususally done using Voigt profile
deblending procedures. Even though blending is a severe limitation
for this analysis at \mbox{$z > 2.5$}, it is interesting to compare the 
observed number of lines with the simulated number of clouds which is 
perfectly defined. 
For \mbox{$N(\mbox{H{\sc i}}) > 10^{14} \mbox{ cm$^{-2}$}$}, 
the lines are not numerous enough for blending to be a problem and their
number is known from observations in the complete range \mbox{$0<z<4$}. 
It can be seen from Fig.~\ref{dndz_all} that 
the evolution of the total number of strong lines is well reproduced.
Data are taken from Lu et al.\ (1991), Petitjean et al.\ (1993)
and Bahcall et al.\ (1993).\par
If the number of lines per unit redshift is approximated by a power-law,
\mbox{${\rm d}n/{\rm d}z \propto (1 + z)^{\gamma}$}, we find 
\mbox{$\gamma \approx 2.6$} for \mbox{$1.5 < z < 3$} and 
\mbox{$\gamma \approx 0.6$} for \mbox{$0 < z < 1.5$}(after smoothing). However the slope
of ${\rm d}n/{\rm d}z$ steepens at \mbox{$z>3$}. Fitting the number density evolution
for \mbox{$1.5 < z < 5$} by a single power law we get 
\mbox{$\gamma \approx 2.9$}. 
The simulated number is also consistent with observation at 
\mbox{$z > 4$} (Williger et al.\ 1994). \par
Fig.~\ref{dndz_sep} shows the contributions of the two populations of clouds 
with \mbox{$N(\mbox{H{\sc i}}) > 10^{14} \mbox{ cm$^{-2}$}$},
$P_{\rm s}$ (solid line) and $P_{\rm u}$ (dash-dotted line). 
It is apparent that the dominant population
is different before and after \mbox{$z \sim 3$}. At high redshift, most of
the lines arise in $P_{\rm u}$ particles whereas at low redshift, most
of the gas is condensed in filamentary structures (see
Petitjean et al.\ 1995).\par
 
Let us assume that most of the $P_{\rm u}$ clouds are co-expanding and optically thin.  If we also assume the flux to be nearly
constant over the redshift range considered, then the evolution of the
column density for each cloud is \mbox{$\propto J_0^{-1} n_{\rm H}^2 l_{\rm c}
\propto (1 + z)^5$}, i.e.\ the column density of such clouds is a rapidly
decreasing function of time. Therefore, at high redshift,  
the number density of clouds situated in the underdense regions 
is, at a given column density, strongly decreasing with time. 
That might explain the very steep slope found for the number
density evolution of the $P_{\rm u}$ clouds as shown 
in Fig.~\ref{dndz_sep}.\par

\begin{figure}
\epsfig{file=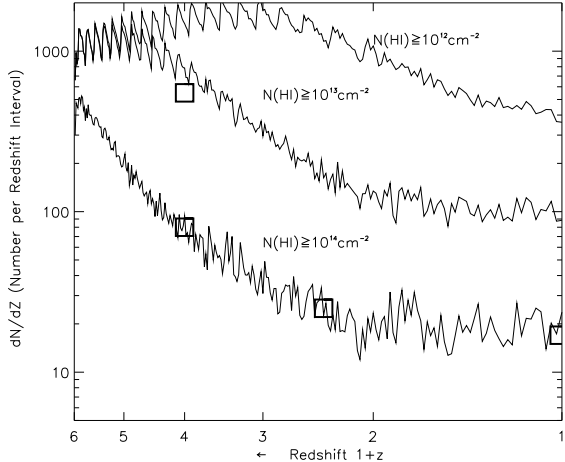,scale=1., width=8cm}
\caption[]{Number density of lines versus redshift for different 
column densities thresholds \mbox{$\log N(\mbox{H{\sc i}})> 12, 13,14$} }
\label{dndz12-14}
\end{figure}

The number density of lines with 
\mbox{$N(\mbox{H{\sc i}}) > 10^{12} \mbox{ cm$^{-2}$}$}
is about constant over the redshift range \mbox{$1<z<5$} 
(see Fig.~\ref{dndz12-14}) and decreases
slowly at lower redshift. Note, however that due to the limited resolution
the number density at \mbox{$z>3$} for column densities 
\mbox{$N(\mbox{H{\sc i}}) <10^{12.5} \mbox{ cm$^{-2}$}$} is probably
underestimated. Low density gas is
found in regions delineated by filamentary structures at high redshift. 
This gas slowly disappears (see Fig.~\ref{3fig}). The total number density 
of lines stays nearly constant because the high column density gas has column
density decreasing with time. This difference in the evolution of the number
density of weak and strong lines has been 
noticed in intermediate resolution data (Bechtold 1994) and confirmed
by Kim et al.\ (1997). The latter authors find \mbox{$\gamma=2.41\pm0.18$} and
\mbox{$1.29\pm0.45$} for \mbox{$\log N(\mbox{H{\sc i}}) >10^{14}
\mbox{ and } 10^{13} \mbox{ cm$^{-2}$}$}
respectively. 

Our model predicts that the number of weak lines should be 
large at low redshift and gives therefore 
an interesting test of the overall picture. Indeed in the line of sight
to 3C273, Morris et al.\ (1991) detect at the 3$\sigma$ level 14 lines with 
\mbox{$w>27 \mbox{ m\AA}$}, only two of them have \mbox{$w>250 \mbox{ m\AA}$} 
and five \mbox{$w>100 \mbox{ m\AA}$}. They deconvolved their spectrum and it can be
seen on their Fig.~1b that a number of very weak lines
are present. 
Although these lines are below the 3$\sigma$ level, this is an indication
that such weak lines are present. Moreover, Tripp et al.\ (1997) have found 
that the number per
unit redshift of lines with \mbox{$w_{\rm r}>50 \mbox{ m\AA}$} is $112\pm21$
at \mbox{$z\sim 0.1$}.
Fig.~\ref{spectrum}
shows a simulated low-redshift spectrum at the resolution of the STIS G140M 
grating on HST with \mbox{${\rm S}/{\rm N}=100$}. It must be noticed that no quick 
conclusion should be drawn from this plot. The wavelength range has
been chosen to illustrate the feasibility of the observations. It can be
seen however that the model predicts a number of very weak features 
corresponding to those seen in the deconvolved spectrum of Morris et al.
The weakest lines have typically \mbox{$w\sim 20 \mbox{ m\AA}$}, corresponding
to \mbox{$N(\mbox{H{\sc i}})\sim 10^{12.3} \mbox{ cm$^{-2}$}$}. \\
\begin{figure} 
\epsfig{file=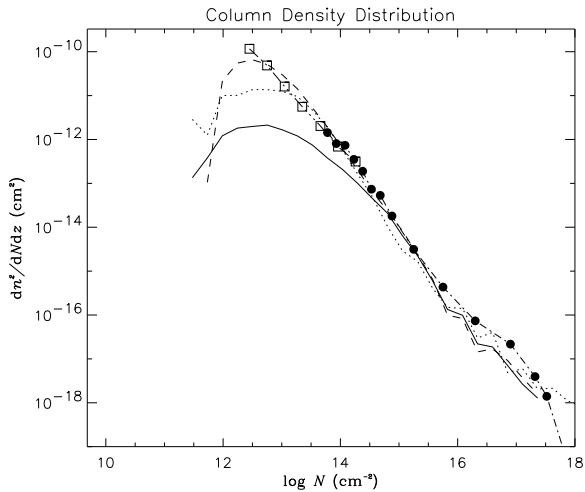 ,scale=1., width=8cm}
\caption[]{H{\sc i} column density distribution at \mbox{$z=3$} computed
directly from the simulation (dashed line ), after reanalysing a sample of
computed spectra using the automatic fitting procedure (dotted line), 
and for population $P_{\rm s}$ only (solid line). 
Observational data points: Hu et al.\ (1996) (square symbols), 
Petitjean et al.\ (1993) (filled circles).}
\label{dndN30_all}
\end{figure}
\begin{figure} 
\epsfig{file=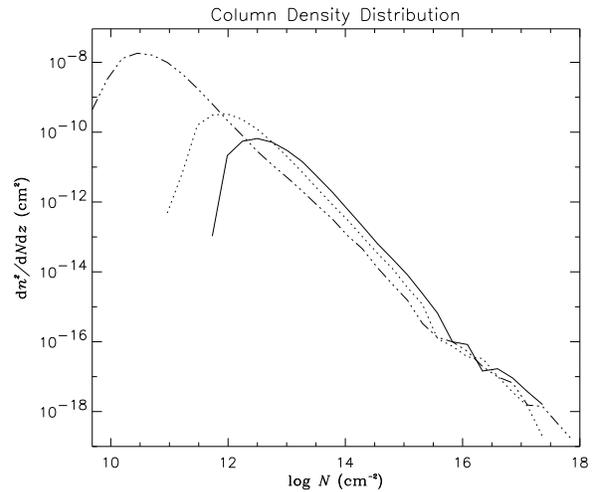,scale=1., width=8cm}
\caption[]{H{\sc i} column density distribution from the simulation at 
redshifts \mbox{$z=0$}, \mbox{$z=2$}, \mbox{$z=3$} (left to right)} 
\label{dndn_zz}
\end{figure}  
\subsection{The column density distribution}\label{coldens}
The number density of lines was computed as in M\"ucket et al.\ (1996)
taking into account that neighboring cells allong the line of sight 
contribute to the same 'cloud'.
The resulting H{\sc i} column density distribution at
\mbox{$z=3$} is given in 
Fig.~\ref{dndN30_all} (dashed line) together with the observed points
taken from Hu et al.\ (1996) (square symbols) and from Petitjean et al.\ (1993)
(diamonds).    
In the simulation, the number of lines with \mbox{$13< \log N(\mbox{H{\sc i}})<14$}
at \mbox{$z=3$} is larger than what is observed by about a factor of
two (see also Zhang et al.\ 1996, Dav\'e et al.\ 1997). 
It is interesting to note that this happens whereas the mean optical 
depth of the Lyman$\alpha$ forest
is well reproduced (see Fig.~\ref{da}). 
Since most of the Lyman forest opacity occurs in clouds with
\mbox{$13< \log N(\mbox{H{\sc i}})<15$} and the differential contribution to
the opacity is at its maximum at \mbox{$\log N(\mbox{H{\sc i}})=14$}
(Kirkman \& Tytler 1997), a larger
number of lines can be found with about the same mean opacity only
if the lines are strongly blended.
This means that, in the simulations, the lines are certainly more clustered 
than in reality. Clustering is a long standing problem and although
very controversal (see Kirkman \& Tytler 1997),
studies using the largest samples available indicate that the 1-D correlation
signal on scales smaller than 200~km~s$^{-1}$ increases with column density 
from 0.1 at \mbox{$\log N(\mbox{H{\sc i}})=13$} to 1 at \mbox{$\log N(\mbox{H{\sc i}})=15$}
(Cristiani et al.\ 1996, Kim et al.\ 1997).

The flattening and turn-over at low column densities is related to 
the resolution of the simulations and depends on redshift 
in agreement with the predictions by M\"ucket et al.\ 1996
(\mbox{\ Eq.~7}): the turn-over happens at larger $N(\mbox{H{\sc i}})$ for higher redshifts
(see Fig.~\ref{dndn_zz}). 
 
The dotted line in Fig.~\ref{dndN30_all} shows the distribution
derived from an averaged sample of synthesized spectra obtained from
the simulation data and fitted using the automatic procedure.
The effect of line blending at low column density is apparent.
The curve is 
in good agreement with the original uncorrected observational 
data for \mbox{$N(\mbox{H{\sc i}})>10^{12.8} \mbox{ cm$^{-2}$}$} (Hu et al.\ 1996).

Finally the solid line in Fig.~\ref{dndN30_all} shows the contribution of 
the particles belonging to the $P_{\rm s}$ population
alone. It is apparent that this population dominates for 
\mbox{$N(\mbox{H{\sc i}}) > 10^{14} \mbox{ cm$^{-2}$}$}. 
It is interesting to note that the column density distribution for 
this population becomes steeper at \mbox{$\log N(\mbox{H{\sc i}})\sim 15$}
and reproduces quite well the departure of the observed function from
a single power-law (Petitjean et al.\ 1993). 
The break seems to be shifted to lower column densities 
at lower redshift (see Fig.~\ref{dndn_zz}) which is at odd with what is 
observed (Kim et al.\ 1997). 
However it is difficult with our simulations to be very specific on
this interesting problem since the statistics in that part of the
distribution is poor. The value at 
\mbox{$\log N(\mbox{H{\sc i}})\sim 16$} should be
more precisely defined both in the observations (but this needs a substantial
number of lines of sight) and in the simulations (but our assumptions
may not be valid for these column densities).
\subsection{The Doppler parameter distribution}\label{doppler}
\begin{figure} 
\epsfig{file=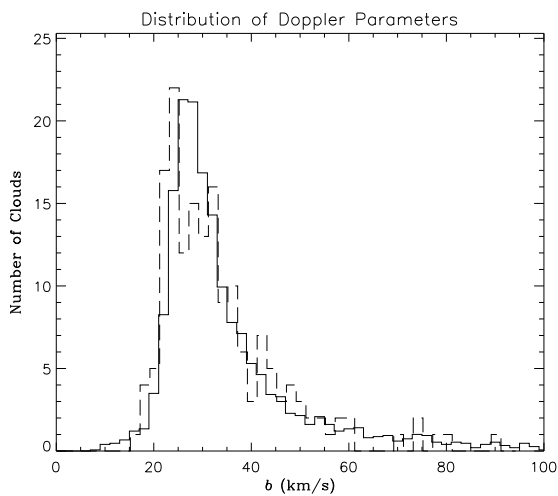,scale=1., width=8cm}
\caption[]{Doppler parameter distribution from line profile fitting of
synthetic spectra (solid line). 
 Keck HIRES results (Hu et al.\ 1996) are overplotted as a 
dashed line}
\label{dndb}
\end{figure}

Fig.~\ref{dndb} shows the
distribution of Doppler parameters obtained from line profile fitting
of synthetic spectra together with the observed distribution
from Hu et al.\ (1996). The large values of $b$ result both from turbulent
motions within the Lyman$\alpha$ complexes (in population $P_{\rm s}$) and
from blending effects. 
Temperature in population $P_{\rm u}$
is restricted to a narrow range around 20\,000~K. Since most of the
weak lines arise from this population, an excess of
$b$ values should be expected at approximately \mbox{20 km s$^{-1}$}. 
However this excess is not
recovered from the synthetic spectra because, as can be seen 
in Fig~\ref{dndN30_all}, most of the weak lines are  
lost due to strong blends. 
A weak dependence of the Doppler parameter distribution with
redshift is seen in the simulations:
The maximum of the distribution is marginally shifted 
to higher Doppler parameters with decreasing redshift consistent
with what is observed (Kim et al.\ 1997).

\begin{figure} 
\epsfig{file=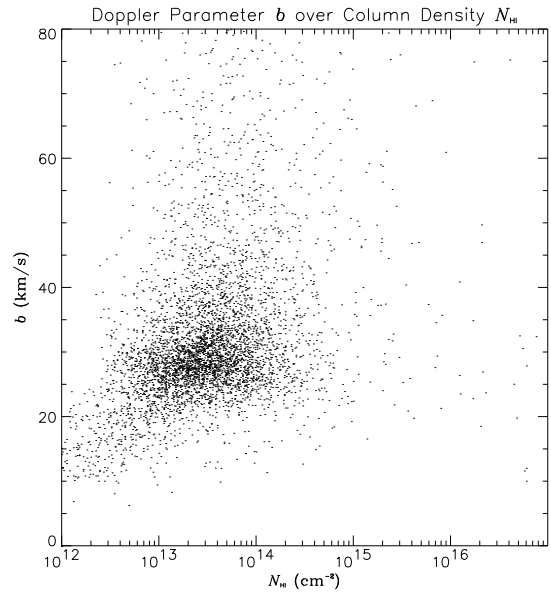,scale=1., width=8cm}
\caption[]{Doppler parameter versus column density from line profile
fitting of synthetic spectra}
\label{dbdn}
\end{figure}

There is a hint for the existence of a lower limit, $b_{\rm c}$, of $b$ that
increases with $N(\mbox{H{\sc i}})$ (see Fig.~\ref{dbdn}). This
has been observed (Kim et al.\ 1997, Kirkman \& Tytler 1997)
and seen in the simulations previously (Zhang et al.\ 1996).
We see this correlation in the outputs of the line profile fitting
of synthetic spectra whereas there is no correlation  
in the simulation, especially the real $b$ distribution is strongly
peaked at \mbox{20 km s$^{-1}$}. 
Although this should be investigated in more detail,
it seems that at least part of the correlation results from blending effects. 
%
%
%
\section{Summary and conclusion}\label{conclusions}
A detailed description of the Lyman$\alpha$ forest has been given by
recent hydro-simulations. The amount of computing time is such however that
most of the descriptions are limited to \mbox{$z>2$}.
Using assumptions that have been shown to be valid in the low density regime
(e.g.\ Bi \& Davidsen 1997) we study the evolution of the Lyman$\alpha$ forest
over the whole redshift range \mbox{$5>z>0$}.

We consider that the gas is photo-ionized and heated by the UV-background.
The evolution of the latter is computed consistently in the simulation assuming
that stars form when the gas cools to temperature smaller than 5\,000~K. 
The normalisation (\mbox{$J_{-21}(z)=0.1$} at \mbox{$z \sim 3$} for
\mbox{$\Omega_{\rm b}h^2$ $\sim 0.0125$})
is obtained by fitting the Lyman$\alpha$ decrement.
This result is consistent with conclusions by others
(Miralda-Escud\'e et al.\ 1996, Weinberg et al.\ 1997, Rauch et al.\ 1997).
At high redshift the Lyman$\alpha$ forest contains most
of the baryons, conclusion achieved previously using independent arguments
(Petitjean et al.\ 1993, Press \& Rybicky 1993, Rauch \& Haehnelt 1995).
Shock-heating is added when the particles are involved in shell-crossing. 
We follow the evolution of two populations of gaseous particles.
Those for which shock heating is large enough to increase the 
gas temperature above the temperature resulting from photo-ionization 
(population $P_{\rm s}$) and the remainder of the particles 
(population $P_{\rm u}$).
Population $P_{\rm s}$ particles are predominantly found within 
dense structures such as filaments and population $P_{\rm u}$ particles 
populate underdense regions. Since stars form predominantly 
in filaments, the gas associated with $P_{\rm s}$ particles is expected to 
contain metals. 

At redshifts \mbox{$z<2.5$} the number density of lines arising from
$P_{\rm u}$ particles 
decreases very rapidly for H{\sc i} column densities larger than
\mbox{$10^{14} \mbox{ cm$^{-2}$}$}. Therefore at low redshifts, the main
contribution to strong Lyman$\alpha$ lines comes from population
$P_{\rm s}$ particles. It is thus not surprizing to observe
that strong Lyman$\alpha$ lines (\mbox{$w_{\rm r}>0.3 \mbox{ \AA}$}) are 
correlated with galaxies at low redshift (Lanzetta et al.\ 1995, 
Le~Brun et al.\ 1996) since galaxies and the Lyman$\alpha$
gas are mostly located in filaments.
 
The number density of lines with \mbox{$\log N(\mbox{H{\sc i}}) > 13$}
at \mbox{$z=0$} is predicted to be \mbox{$\sim 100$} consistent with recent
observations by Tripp et al.\ (1997).
The number density of lines with \mbox{$\log N(\mbox{H{\sc i}}) > 12$}
remains about constant over the redshift range \mbox{$3>z>1$} and decreases
slowly at lower redshift. This is the result of both
decreasing ionizing flux and decreasing mean hydrogen density. 
The number density
of such weak lines is predicted to be of the order of 400 per unit redshift
at \mbox{$z\sim0$}. This prediction can be tested along the line of sight
to 3C273 with the new instrumentation on the Hubble Space Telescope.

\acknowledgements
We wish to thank Martin Haehnelt for his comments and suggestions.
R.~Riediger was supported by a fellowship (Mu 1043/3-1) from the DFG Germany.
\endacknowledgements

\end{document}